\documentstyle[12pt,aaspp4]{article}
\begin{document}

\title{Gravitational Microlensing with the Space Interferometry Mission}

\author{Bohdan Paczy\'nski}
\affil{Princeton University Observatory, Princeton, NJ 08544--1001, USA}
\affil{e-mail: bp@astro.princeton.edu}

\begin{abstract}

The Space Interferometry Mission (SIM), with its launch date planned
for 2005, has as its goal astrometry with $ \sim 1 ~ \mu $ arcsecond
accuracy for stars as faint as 20th mag.  If the SIM lives to expectations
it can be used to measure astrometric displacements in the light
centroid caused by gravitational microlensing in the events
detected photometrically from the ground.  The effect is
typically $ \sim 0.1 $ mas, i.e. two orders of magnitude larger than
planned SIM's accuracy.  Therefore, it will be possible
to determine the mass, the distance, and the proper motion of
almost any MACHO capable of inducing a photometric microlensing event
towards the galactic bulge or the Magellanic Clouds, even though no light 
from the MACHO has to be detected.

For strong microlensing events in which the source is 
photometrically resolved, like the recent MACHO 95-30 event,
SIM's astrometry combined with accurate ground based photometry
will allow the determination of the angular stellar radii, and
therefore the effective temperature of the source.

The effective astrometric cross sections for gravitational lensing 
by nearby high proper motion stars and brown dwarfs are $ \sim (1'')^2 $ 
and the effective time scales are $ \sim 1 $ year.  SIM will provide 
the only practical way to measure masses of single nearby objects
with $ \sim 1\% $ accuracy.  The times of lensing events can be
predicted years ahead of time.

\end{abstract}


\section{Introduction}

The searches for gravitational microlensing events in the galactic bulge
and in the Magellanic Clouds have matured, and frequent alerts of new events 
are provided in real time (cf. Paczy\'nski 1996a for a review).
Unfortunately, currently there is no way to firmly determine the distance to
a lensing object and to measure its mass because a number of physical
parameters combine into a single observable quantity: the time scale $ t_0 $.
In some cases the degeneracy may be partly broken, as reviewed by Gould (1996).
Also, at least one suggestion was made how to remove the degeneracy 
altogether in some rare cases (extreme microlensing: Gould 1997).  

H${\rm \o}$g, Novikov \& Polnarev (1995), Miyamoto \& Yoshi (1995), and Walker 
(1995) pointed out that accurate astrometry could permit the determination of
the distance and the mass of a MACHO.  It seems that the Space 
Interferometry Mission (SIM) is
the first specific instrument which will have the capability adequate
for this task.  According to the mission specification,
it will have an angular resolution of
$ \sim 10 $ mas, and positional accuracy down to $ \sim 1 $ micro-arcsecond,
for stars as faint as 20 mag.  A complete description of the technical
details as well as the scientific program were published by
Allen, Shao, \& Peterson (1997), and can also be found on the World
Wide Web at:
\centerline{\it http://huey/jpl.nasa.gov/sim/ }
\noindent
Some of the topics covered in this letter (section 3) are presented in
much more details in the paper written by three members of 
the SIM team: Boden, Shao, \& Van Buren (1997).  

Recently, a somewhat related aspects of gravitational lensing
astrometry were considered by Miralda-Escud\'e (1996) and Paczy\'nski (1996b),
who in particular pointed out that astrometric cross section is much
larger than photometric cross section.  Therefore, the masses of single
high proper motion stars and brown dwarfs can be determined accurately
by measuring the astrometric lensing effects of the distant stars.

The purpose of this paper is to present in one place various suggestions 
made in the past for the application of $ \sim 1 $ micro-arcsecond astrometry
to determine the masses of stars, brown dwarfs, and MACHOs, whatever the
MACHOs might be.  In addition, a new suggestion to use such astrometry
to measure stellar radii and effective temperatures is also presented.


\section{Astrometry of microlensing}

Typical angular separations of multiple images formed by microlensing
within our galaxy or towards the Magellanic Clouds are $ \sim 1 $ mas
(Paczy\'nski 1996a, and references therein), so they will remain
unresolved by the SIM.  However, the centroid of the combined image
is expected to move around by $ \sim 1 $ mas, and this motion can be
readily measured by the SIM with a very high precision. 

The angular Einstein ring radius of a point lens is given as
\begin{equation}
\varphi _E = \left[ \left( { 4GM \over c^2 } \right) 
\left( { D_s - D_d \over D_sD_d } \right) \right] ^{1/2} =
 0.902 ~ {\rm mas} ~ \left( { M \over M_{\odot} } \right) ^{1/2}
\left[ 10 ~ {\rm kpc} \times \left( { 1 \over D_d } 
 - { 1 \over D_s } \right) \right] ^{1/2} ,
\end{equation}
where $ D_s $, $ D_d $, and $ M $ are the distances to the source and
the lens (deflector), and the lens mass, respectively.  The time scale
of the event is defined as
\begin{equation}
t_0 \equiv { \varphi _E \over \dot \varphi } ,
\end{equation}
where $ \dot \varphi $ is the proper motion of the lens relative to
the lensed star (note that the MACHO collaboration uses $ 2 t_0 $ for
the time scale).  

In the simplest case the relative motion of the lens and the source
is linear, with the angular impact parameter $ \varphi _{min} $.
Superposed on this is the relative parallactic motion with the
angular amplitude $ \pi _{ds} $ (in radians):
\begin{equation}
\pi _{ds} =  1 ~ {\rm AU} \times \left( { 1 \over D_d } - { 1 \over D_s } 
\right) , \hskip 1.0cm  1 ~ {\rm AU = 1.5 \times 10^{13} ~ cm } ,
\end{equation}
where the astronomical unit (AU) is the radius of Earth's orbit.
The equations (1) and (3) may be combined to obtain the lens mass:
\begin{equation}
M = 0.123 ~ M_{\odot} ~ { \varphi _E^2 \over \pi _{ds}} ,
\end{equation}
where the Einstein ring radius and the relative parallax of the lens
with respect to the source are expressed in milli-arcseconds.

The two micro images formed by the lens cannot be resolved, but the 
light centroid is displaced with respect to the source position by the angle
(cf. H${\rm \o}$g, Novikov \& Polnarev 1995, Miyamoto \& Yoshi 1995, 
Walker 1995)
\begin{equation}
\delta \varphi = { \Delta \varphi \over u^2 +2 } ,
\hskip 1.0cm u \equiv { \Delta \varphi \over \varphi _ E } ,
\end{equation}
where $ \Delta \varphi $ is the angle between the source and the lens,
and the lens is assumed to be dark.  The maximum displacement is
\begin{equation}
\delta \varphi _{max} = 8^{-1/2} \varphi _E \approx 0.354 \varphi _E ,
\hskip 1.0cm {\rm for} \hskip 1.0cm u = 2^{1/2} .
\end{equation}

For $ u \gg 1 $ the light centroid corresponds almost exactly to the
position of the dominant image (cf. Paczy\'nski 1996b).
Note, that if the lensing star is invisible, as assumed, then the source
position can be measured with respect to any nearby star.

\section{Astrometric microlensing in the LMC}

In this section we present a few
examples of the displacement trajectories to illustrate the principle
of the phenomenon, and to demonstrate that the lens mass and parallax
have distinct effects which should be readily measurable with the SIM.

Let us consider a microlensing event with a time scale $ t_0 = $ 50 days
and maximum magnification $ A_{max} = 5.07 $, which corresponds to the
impact parameter $ u_{min} = 0.2 $.  We adopt the source distance to be
$ D_s = 50 $ kpc, and consider three different
cases of the lens distance: 10, 30, and 45 kpc,
with identical lens mass of $ 0.3 ~ {\rm M_{\odot} } $.  The corresponding
Einstein ring radii are $ \varphi _E = 0.44, ~ 0.18, ~ 0.074 $ mas,
respectively.  Finally, the relative proper motion is 
$ \dot \varphi = \varphi _E / t_0 = 3.2, ~ 1.32, ~ 0.54 ~ 
{\rm mas ~ yr^{-1} } $, for these three cases. 

For our example we adopt $ \sin \beta = - 0.99 $, which is reasonable
for a star in the LMC, and we arbitrarily take $ \dot \varphi _{ \lambda } = 
2 \dot \varphi _{ \beta } $, where ($ \lambda , \beta $) are the ecliptic
longitude and latitude, respectively.
The relative parallax is equal $ \pi _{ds} = 0.080, ~ 0.013, ~ 0.0022 $ mas
for the three cases, respectively.  Finally, we chose
the maximum magnification and the smallest
angular separation between the lens and the source to be at the time
when the star is at the largest angular separation from the sun.

The three examples of astrometric effects of microlensing a star in the
LMC are presented in Figure 1, which shows the relation between the
displacements in the two ecliptic coordinates, ($ \lambda , \beta $).
All lines in Fig.1 begin 18 months before the
maximum of photometric magnification, and they end 18 months after
that maximum.  Both ends are close to
$ ( \delta  \lambda , \delta  \beta  ) = (0,0) $.
The largest curve corresponds to the lens
distance $ D_d = 10 $ kpc, the smallest to $ D_d = 45 $ kpc.
Note, that the largest displacement is proportional to the Einstein
ring radius, with $ \varphi _E = 8^{1/2} \delta \varphi _{max} $
(cf. eq. 6).  The relative proper
motion of the lens with respect to the star is given as the ratio:
$ \Delta \dot \varphi = \varphi _E / t_0 $.  

\begin{figure}[t]
\plotfiddle{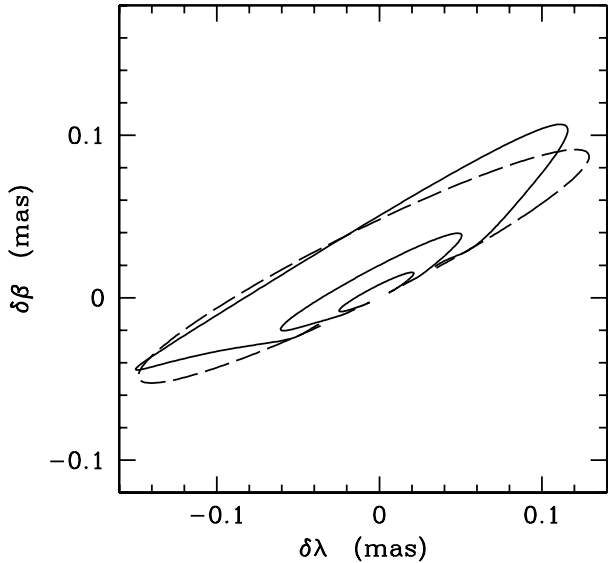}{8cm}{0}{50}{50}{-160}{-80}
\caption{
The three solid curves show the astrometric displacement in ecliptic 
coordinates ($ \lambda , \beta$) caused by three microlensing events 
described in the text.  The source is a star in the LMC at the distance 
$ D_s = 50 $ kpc, and the lenses of $ 0.3 ~ M_{\odot} $ are located at 
the distances $ D_d = $ 10, 30, and 45 kpc.  The larger the distance the 
smaller the displacement.  In all three cases the microlensing event had
the time scale $ t_0 = $ 50 days, and the impact parameter $ u_{min} = 0.2 $,
corresponding to the maximum magnification $ A_{max} = 5.07 $.
The largest displacement of a given
curve from the origin is $ \sqrt{8} $ times smaller than the corresponding
Einstein ring radius.
The dashed curve corresponds to the $ D_d = 10 $ kpc case with the
effect of earth orbital motion artificially suppressed.
}
\end{figure}

\begin{figure}[t]
\plotfiddle{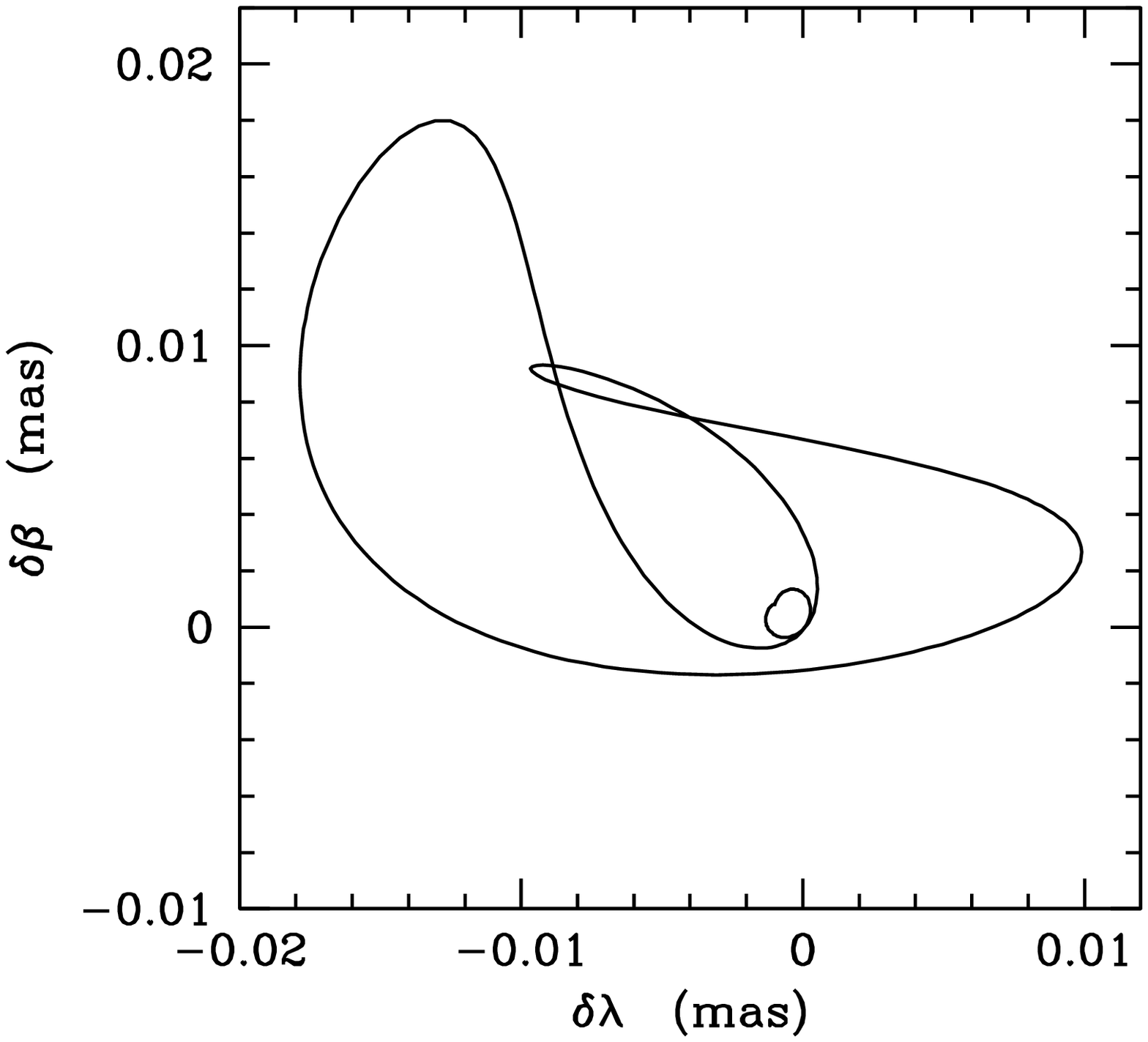}{8cm}{0}{50}{50}{-160}{-80}
\caption{
The difference between the solid and the dashed curve corresponding to
$ D_d = 10 $ kpc (the pair of largest curves) in Figure 1 is shown in the
same coordinate system.  This demonstrates the scale of the parallactic
effect.
}
\end{figure}

The determination of the relative parallax requires model fitting,
but this should be feasible.  The dashed curve in Figure 1 shows
the displacement of the image centroid for the case when the lens
is at $ D_d = 10 $ kpc, but the earth orbital motion is artificially
suppressed.  The difference between the solid and the dashed lines
is shown in Figure 2.  This is the complicated effect of the earth
motion, and its scale is such that the SIM should be able to measure it
readily.  Therefore, if the MACHOs responsible for microlensing
of the LMC stars are located in the galactic halo then their
distances and masses will be determined with the SIM astrometry.
The masses and distances to the galactic bulge lenses will be
even easier to measure as their parallaxes are relatively large.

Note, that $ \sim 1 ~ {\rm \mu s } $ astrometry will provide
accurate astrometric determination of the MACHO mass, as shown with
the scale of the Figure 2,   However, a much lower precision of
the ROEMER mission (cf. H${\rm \o}$g, Novikov \& Polnarev 1995), or of the
GAIA project (Lindegren \& Perryman 1996) will not provide a useful MACHO
mass determination.

There is a region of parameter space in which the effects of
acceleration in earth's orbital motion are too small for the SIM
to measure (cf. Boden et al. 1997).  This includes 
very short events with $ t_0 \ll 1 $ year, and the case when the relative
parallax $ \pi _{ds} $ is very small.  So, if the lenses responsible
for the LMC lensing are in the LMC itself (Sahu 1994) then the
astrometric effects are so small (cf. Figure 1) that it will be
impossible to determine their masses and distances accurately.
However, it will be still possible to establish that they are
close to the sources, i.e. within the LMC rather than in the
galactic halo.

\section{Astrometric microlensing of high proper motion stars}

All classical microlensing searches
monitor millions of stars to detect a few events which come at random time.  
Miralda-Escud\'e (1996) and Paczy\'nski (1995, 1996b) pointed out that
there is a very different regime of microlensing when we begin not
with the sources but with the lenses.  This works best when the lenses are
selected as stars with very large proper motions, indicating they
are relatively nearby, and therefore their angular Einstein ring radii 
are relatively large (cf. equation 1).

The astrometric effects have a large effective cross-section.
For example, the Barnard's star (van de Kamp 1971) has a 
parallax of $ 0.''522 $, and a proper motion of $ 10.''31 $ per year.
If we adopt its mass to be $ M_B \approx 0.2 ~ {\rm M_{\odot}} $, we 
obtain for its Einstein ring radius
\begin{equation}
\varphi _{EB} = 30 ~ {\rm mas} \times 
\left( { M_B \over M_{\odot} } \right) ^{1/2} .
\end{equation}
A position of a  distant background star passing by at an angular distance
of $ 9.''0 $ is
displaced by lensing by 100 $ \mu {\rm s }$ -- a huge effect for the 
SIM, which in this case could determine the mass of Barnard's star
with $ \sim 1\% $ accuracy.  
The large proper motion combined with the large astrometric cross section
implies that the useful background stars may be looked for in a large
area in the sky.  Adoptic 3 years as a pessimistic life span for SIM
implies that the Barnatd's star alone will cover an area
$ \sim 31'' \times 18'' = 0.15 \times (1')^2 $ in the sky.  

A lower limit to the number of lensing events for which the SIM will be able
to make accurate mass determination for the lenses can be obtained using
Hipparcos catalog (Perryman et al. 1997).  There are 12,204 stars in the
catalog with the distance less than 100 pc and proper motion $ \dot \varphi $
larger 
than $ 100 ~ {\rm mas ~ yr^{-1}} $.  Their masses can be estimated with the
mass - luminosity relation, and therefore the corresponding Einstein ring
radii can be calculated.  Adopting $ 100 ~ {\rm \mu s} $ as the 
required astrometric
displacement $ \delta \varphi $ due to lensing, the minimum impact parameter
$ \Delta \varphi $ can be calculated with eq. 5.  Combining all these data 
it is possible to calculate the time scale for lensing event, $ t = 
\Delta \varphi / \dot \varphi $, and the area of the sky swept by each star 
during a three year time interval: $ a = 3 \dot \varphi
 \times 2 \Delta \varphi $.  
The total area swept by all 12,204 stars is $ 12.4 \times (1')^2 $, i.e.
$ \sim 10^{-7} $ of the whole sky.  As stars within 100 pc are distributed
isotropically in the sky,
we may expect to find $ \sim 18 $ stars brighter than $ V = 17 $,
mag and $ \sim 65 $ stars brighter than $ V = 20 $ mag (cf. Table 4.2
of Mihalas \& Binney 1983) in the total area swept by these stars during 
three years.  A median time scale for the microlensing events,
weighted by the area in the sky swept by the stars, turns out
to be a convenient 7 months, with 80\% of event time scales expected 
in the range 2 months - 2 years.  For each of the events a $ 1 ~ {\rm \mu s }$
astrometric accuracy implies 1\% accurate mass determination.

Note, that in this case the lensing star is bright, and the lensed source
is faint, but clearly resolved with the SIM.  The relative astrometry of
the two stars will provide the relative proper motion $ \Delta \dot \varphi $,
the relative parallax $ \pi _{ds} $, and the Einstein ring radius 
$ \varphi _E $ from the analysis of displacements $ \delta \varphi $
as given with eq. (5), which in this case refers to
the dominant image of the source.  The lens mass follows from eq. (4).

If the gravitational displacement requirement is reduced from 
$ 100 ~ {\rm \mu s} $ to $ 30 ~ {\rm \mu s} $, then the total 
area swept by the 12,204 high proper motion Hipparcos stars in 3 years
increases to $ 41.4 \times (1')^2 $, i.e. it is larger by a factor
3.3 compared to the previous case, and the number of possible
events increases by the same factor.  A median time scale increases
to 1.6 years.  This shows the importance of a precise astrometry:
the higher the accuracy, the more lensing events can be used
for accurate mass determination.

Obviously, there are many more stars which are too faint to be in the
Hipparcos catalog, but which have large proper motion (e.g. Luyten 1976),
and will be suitable for accurate mass determination.  
It would be very interesting to have the entire sky searched for faint,
(20th mag) high proper motion ($ > $ 1''/year) stars, as some of them may 
turn out to be brown dwarfs,  The difference between a star and a
brown dwarf is best established with an accurate mass determination,
and this will be possible with the SIM.

\section{Effective temperature scale with astrometric microlensing}

When the angular impact parameter is smaller than the angular source
radius then an accurate microlensing light curve can be used to obtain
the ratio of the two angles.  If the effective temperature of the source
is estimated from its color and/or spectrum then its surface
brightness can also be estimated; combining this with the observed
flux can be used to determine the source angular diameter, and
the angular Einstein ring radius.  Finally, the relative proper motion 
is obtained by dividing the Einstein ring radius by the event time scale,
hence the name: `proper motion event' (cf. Gould 1996, and references therein).

It may be useful to reverse this process as
stellar effective temperatures are not known all that well.
SIM's astrometry of microlensing events will provide accurate measurements
of the relative proper motions between the lens and the source.
Combining this with an accurate light curve of a
`proper motion event' will give 
the angular stellar radius, which in turn will be used to determine the
effective temperature of the lensed star.  

The first spectacular example of a microlensing event of this type was
recently provided by the MACHO Alert 95-30 (Alcock et al. 1997). 
The ratio of the angular impact parameter to the angular source radius was
determined to be $ 0.715 \pm 0.003 $.
Had the Einstein ring radius been measured with SIM-like astrometry
the angular diameter of the lensed star would have been known to
$ \sim 1\% $ accuracy, comparable to the best existing 
stellar measurements.  Of course, accurate astrometry was not available,
and the source colors and spectra were used to estimate stellar radius, and
the proper motion with a precision of only $ \sim 20\% $.

The example of MACHO Alert 95-30 demonstrates that the accuracy of 
stellar radii determination with $ \sim 1 ~ {\rm \mu s} $
astrometry of resolved microlensing
events is comparable to that achieved with the best 
detached eclipsing binaries (cf. Andersen 1991), or the
best interferometric measurements (Shao \& Colavita 1992).
Once the angular stellar radius is measured, the determination of the
effective temperature is a standard procedure.

\section{Summary} 

A $ \sim 1 ~ {\rm \mu s} $ astrometry of microlensing events can be
used to determine the lens masses and the source radii (and hence
their effective temperatures), but it requires Target of Opportunity mode
of operation.  In order to be useful the Space Interferometry
Mission (Allen, Shao,\& Peterson 1997, Boden, Shao \& Van Buren 1997)
would have to respond to ground based alerts within a few weeks.

The only way to measure masses of single objects with an accuracy
of $ \sim 1\% $ is with gravitational lensing.  Lensing events
caused by nearby high proper motion stars passing within $ 1''-10'' $ of
distant background stars, can be predicted many years into the future,
making it possible to prepare the SIM's observing program ahead of time,
with no need for the TOO operation.  

\acknowledgments{It is a great pleasure to thank Dr. D. Peterson and
the anonymous referee for many useful comments.
This work was supported with the NSF grants AST--9313620 and AST--9530478.}  



\end{document}